\documentclass{article}[12pt]
\usepackage{amssymb}
\usepackage{fancyhdr}
\usepackage{color}

\usepackage{natbib}
\usepackage{graphics,graphicx}

\def\checkbox{\square\!\!\!\!\!\!\!{\bf X}}

\usepackage{color}

\begin{document}
\thispagestyle{empty}
\begin{center}

{\large\bf Astro2020 Science White Paper}

\bigskip{\Large\bf AGN (and other) astrophysics with Gravitational Wave Events}

\end{center}

\vfill

\noindent{\bf Thematic Areas:}
$$\vbox{\hsize 5truein \halign{# \hfil & #\hfil \cr
$\square$ Planetary Systems & $\square$ Star and Planet Formation \cr
$\checkbox$ Formation and Evolution of Compact Objects & $\square$ Cosmology and Fundamental Physics \cr
$\square$  Stars and Stellar Evolution & $\square$ Resolved Stellar Populations and their Environments \cr
$\checkbox$ Galaxy Evolution & $\checkbox$ Multi-Messenger Astronomy and Astrophysics \cr
}}$$

\vfill

\begin{center}
{\it Authors:}

\vskip 3mm
{K. E. Saavik Ford (CUNY Borough of Manhattan Community College/Am. Museum of Natural History), Imre Bartos (U of Florida), Barry McKernan (CUNY Borough of Mahnattan Community College/Am. Museum of Natural History), Zoltan Haiman (Columbia University), Alessandra Corsi (Texas Tech University), Azadeh Keivani (Columbia University), Szabolcs Marka (Columbia University), Rosalba Perna (SUNY Stony Brook/CCA), Matthew Graham (CalTech), Nicholas P. Ross (U. of Edinburgh), Daniel Stern (CalTech/JPL), Jillian Bellovary (CUNY Queensborough Community College/Am. Museum of Natural History), Emanuele Berti (JHU), Matthew O'Dowd (CUNY Lehman College/Am. Museum of Natural History), Wladimir Lyra (CSUN),
Mordecai-Mark Mac Low (Am. Museum of Natural History/CCA), Zsuzsanna Marka, (Columbia University)  \\ 

\bigskip{\it Endorsed by:}

Brian D. Metzger (Columbia University), Filippo D'Ammando (INAF), Brian Humensky (Columbia University), Richard O'Shaughnessy (RIT), Peter Meszaros (Penn State), Nathan W. C. Leigh (Universidad de Concepci\'on/AMNH), Margarita Karovska (CfA), Peter Shawhan (University of Maryland and Joint Space-Science Institute), Steven L. Liebling (Long Island University), Paolo Coppi (Yale) \\}
\end{center}
\setlength{\parindent}{0.25in}
\noindent
First author contact information: K. E. Saavik Ford\\
Dept. of Astrophysics, American Museum of Natural History, Central Park West at 79th St., New York, NY 10024.\\
Physics Program, Graduate Center of the City University of New York, 365 5th Avenue, New York, NY 10016.\\
Dept. of Science, CUNY BMCC, 199 Chambers St., New York, NY 10007.\\
Email: sford@amnh.org\\

\newpage
\begin{abstract}
The stellar mass binary black hole (sBBH) mergers presently detected by LIGO may originate wholly or in part from binary 
     black hole
mergers embedded in disks of gas around supermassive black holes. Determining the contribution of these active galactic nucleus (AGN) disks to the sBBH merger rate enables us to uniquely measure important parameters of AGN disks, including their typical density, aspect ratio, and lifetime, thereby putting unique limits on an important element of galaxy formation. For the first time, 
gravitational waves will allow us to reveal the properties of the hidden interior of AGN disks, while 
    electromagnetic radiation (EM)
probes the disk photosphere.
The localization of sBBH merger events from LIGO is generally insufficient for association with a single EM counterpart. However, the contribution to the LIGO event rate from rare source types (such as AGNs) can be determined on a statistical basis. To determine the contribution to the sBBH rate from AGNs in the next decade  requires: {\em 1) a complete galaxy catalog for the LIGO search volume, 2) strategic multi-wavelength EM follow-up of LIGO events and 3) significant advances in theoretical 
     understanding of AGN disks and the behavior of objects embedded within them.
     }

\end{abstract}

    \section{Is LIGO detecting the mergers of black holes embedded in active galactic nucleus disks?}

LIGO has detected several stellar mass black hole binary (sBBH) mergers in gravitational waves (GW). It remains unclear if there are counterparts to these events, detectable with other messengers, but there may be (see the ``Multi-Messenger Astrophysics Opportunities with Stellar-Mass Binary Black Hole Mergers'' whitepaper). 
The GW events detected by LIGO may result from sBBH mergers in the disks of active galactic nuclei (AGN) \citep[Fig.~1;][]{McKF12,McKF14,2017ApJ...835..165B,2017MNRAS.464..946S,McKernanFordetal2018,Fragione2018}.  

Multimessenger astronomy can be used to investigate such events over the coming decade, even without detectable counterparts. First, we can infer types of electromagnetic radiation (EM) counterparts on a statistical basis \citep{Bartosetal2017}. Second, we can uniquely constrain active AGNs and nuclear star clusters (NSCs) 
    as sources, 
by comparing AGN and NSC models with observations of rates, masses and spin distributions for sBBH mergers \citep[e.g.][]{McKernanFordetal2018,Fordinprep,MckFOWinprep}. 

\vspace{3mm}  
   \noindent\begin{minipage}{0.4\textwidth}
 
{ { Figure \ref{fig:cartoon}.  Cartoon of a swarm of stellar mass BHs in a galactic nucleus embedded in and orbiting through an AGN disk around a supermassive BH. Some fraction of BHs in the galactic nucleus will end up embedded in the disk, 
    where 
gas torques lead to migration and mergers of BHs detectable with LIGO \citep{McKF14,Bellovary16,Bartosetal2017,2017MNRAS.464..946S} \label{fig:cartoon}
 }
}
\end{minipage}%
\hfill%
\noindent\begin{minipage}{0.6\textwidth} 
\includegraphics[width=0.8\textwidth]{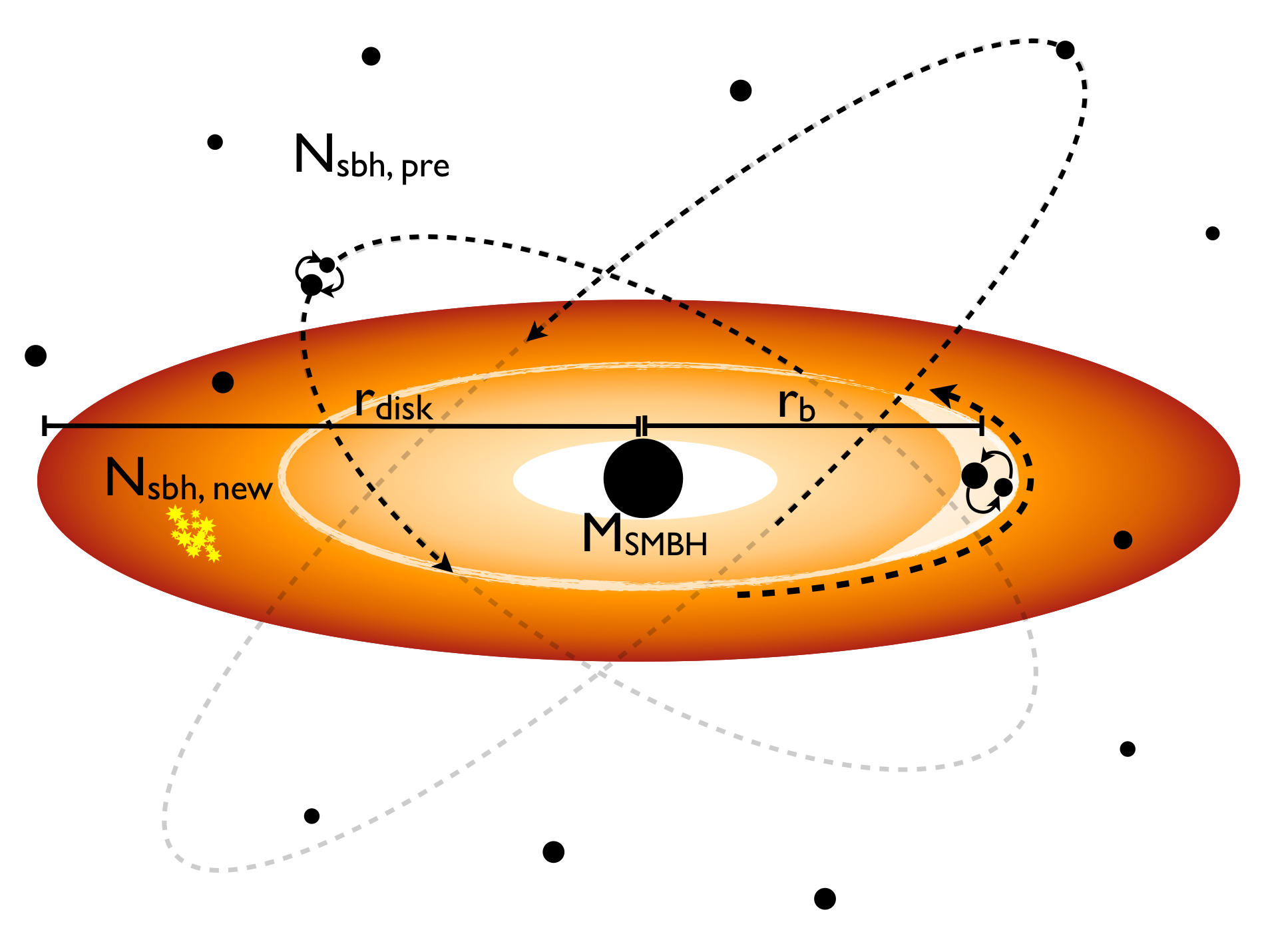}
\vspace{3mm}
\end{minipage}

\section{How do we assign GW events to AGNs?}
\label{sec:assign}
GW localization alone cannot pinpoint a single host galaxy. sBBH mergers can be localized with LIGO to a comoving volume of $V_{\rm GW} \sim 10^4$--$10^9$\,Mpc$^{3}$ at 90\% confidence level \citep{2016arXiv161201471C}. However, given enough events, we can constrain the contribution of a sub-population of galaxies (e.g. AGNs) to the GW event-rate. If we restrict sBBH mergers to the brightest AGNs, the  number density is  $n_{\rm AGN}=10^{-4}$--$10^{-5}$\,Mpc$^{-3}$. For such rare source types, correlation with GW localizations can be established within a few years of operation by Advanced LIGO and Advanced Virgo, even if only a fraction of sBBH mergers occur in AGNs \citep{Bartosetal2017}. 

Assume that there are two sBBH formation channels: one associated exclusively with AGNs (given by fraction $d_{\rm AGN}$\footnote{In principle, LIGO events can be spatially correlated with AGNs even if they are unrelated to AGNs, but occur in galaxies whose spatial distribution is correlated with AGN. The cross-correlation length between local galaxies and quasars is $\sim 6$\,Mpc, \cite{Shen+2013} which is an order of magnitude smaller than the linear size of the typical LIGO error volume \cite{2016arXiv161201471C}. This effect is therefore small, unless the events occur in rare galaxy sub-types that have a stronger correlation with AGN.}), and one associated with all galaxy types, much more numerous than AGN. To identify an AGN contribution, we can require a $3\sigma$ excess over the expectation for no AGN contribution, and set a required confidence interval of 95\% certainty. \cite{Bartosetal2017} show that these criteria determine the required number of GW merger detections, $N_{\rm GW}$, depending only on $d_{\rm AGN},n_{\rm AGN}$, and $V_{\rm GW})$.

For example, if $d_{\rm AGN}=1$, i.e. all events come from AGN, the required number of detections to identify an AGN population is only $N_{\rm GW} \sim100$. This number of observed mergers is expected to be reached after a few years of operation with Advanced LIGO and Advanced Virgo at design sensitivity, or possibly sooner once LIGO A+ occurs. If $d_{\rm AGN}=0.3$, we find that we need $N_{\rm GW} \sim 600$ to statistically establish the presence of an AGN contribution \citep{Bartosetal2017}. However, for a contribution of $d_{\rm AGN}\lesssim 0.1$ to the observed LIGO rate, this technique by itself will have difficulty establishing the presence of an AGN sub-population and other messengers or techniques are required.

The above assumes perfectly complete and uncontaminated galaxy catalogs. GWs from sBBHs can be detected out to a few gigaparsecs with Advanced LIGO and Advanced Virgo at design sensitivity \citep{2018LRR....21....3A}. {\textbf{No galaxy catalog is even close to complete at this distance range.}} For incomplete catalogs, \cite{Bartosetal2017} show that the required detection number $N_{\rm GW}$ is inversely proportional to the completeness of the AGN catalog used for the search. The effect of contamination can be assessed similarly. If $f_{\rm cont}$ is the fraction of objects in the catalog that are misidentified as AGN, then assuming $n_{\rm AGN}V_{\rm GW}\gg 0.09d_{\rm AGN}$, $N_{\rm GW}\propto(1-f_{\rm cont})^{-1}$. For our example above with $d_{\rm AGN}=1$, if $f_{\rm complete}=0.78$ and $f_{\rm cont}=0.05$, $N_{\rm GW}$ increases from $\sim$100 to $\sim$200.{\textbf{ We need a complete AGN catalog for the LIGO GW search volume, with low contamination by false positives, in order to successfully constrain the fraction of mergers coming from AGN}}.

\section{What can we infer from the LIGO-Virgo GW rates?}
\label{sec:inference}
We have already learned something about AGNs and NSCs from current LIGO-Virgo measurements. We can parameterize the expected rate of sBBH mergers in AGN as \citep{McKernanFordetal2018}
\begin{equation}
{\cal R}=12 \mbox{ Gpc}^{-3} \mbox{ yr}^{-1}\frac{N_{\rm GN}}{0.006 \mbox{ Mpc}^{-3}} \frac{N_{\rm BH}}{2 \times 10^{4}} \frac{f_{\rm AGN}}{0.1}
\frac{f_{d}}{0.1}\frac{f_{b}}{0.1} \frac{\epsilon}{1}\left(\frac {\tau_{\rm AGN}}{10\mbox{ Myr}}\right)^{-1}
\label{eq:rate1}
\end{equation}
where $N_{\rm GN}$ is the number density of galactic nuclei, $N_{\rm BH}$ is the number of sBHs in the central parsec of a galactic nucleus, $f_{\rm AGN}$ is the fraction of galactic nuclei that are AGN, $f_{d}$ is the fraction of BH from the NSC that are embedded in the disk, $f_{b}$ is the fraction of those BH in the disk that are in binaries and $\tau_{\rm AGN}$ is the AGN disk lifetime. For a simple gas disk of constant aspect ratio, $h$, and ignoring orbital grind-down, 
$f_{d}=3h$. Present best estimates of AGN and NSC parameters are sufficiently uncertain that a rate of ${\cal R} \sim 10^4$--$10^{-4}$ sBBH mergers Gpc$^{-3}$ yr$^{-1}$ is allowed \citep{McKernanFordetal2018}! LIGO-Virgo measurements set a firm upper limit on the rate of $108$ Gpc$^{-3}$ yr$^{-1}$ at $90\%$ confidence \citep{LatestLIGORatePaper}. So already {\textbf{we can constrain important parameters of AGN disks and the typical number of stellar mass black holes found in NSCs, using GW measurements alone}}. Independent constraints on any one of these parameters using multiple messengers, will allow us to break degeneracies and measure the remaining parameters to even higher precision. 

Some $\sim 1/3$ of all galactic nuclei in the local Universe display  nuclear activity \citep{Ho2008}, ranging from the most common  low-ionization emission regions (LINERs) to $\sim 10\%$ of galaxies containing Seyfert nuclei and $<1\%$ quasars. If we assume that all low luminosity AGNs, including LINERs, are actually optically thick radiatively inefficient accretion flows (RIAFs), then we end up with a very large number of thick disks that can trap and merge a large number of sBH.  Assuming then $f_{AGN} \sim 0.3, f_{b} \sim 0.1$ and $N_{\rm sBH} \sim 2 \times 10^{4}$ in Eqn.~(\ref{eq:rate1}), Fig.~2 shows the LIGO upper and lower merger rate bounds (diagonal lines) on typical 
$h$ and $\tau_{\rm AGN}$ 
\citep{Fordinprep}. Such RIAFs must have large typical aspect ratios \citep[$h>0.5$; e.g.][]{Narayanetal1997}. Thus LIGO is already telling us that most {\textbf{LINERs cannot be optically thick RIAFs}} \citep{McKernanFordetal2018}. If other methods or messengers can eliminate the possibility that AGN driven mergers contribute significantly to the LIGO measured rate, we can restrict AGN disk parameters to lie below the lower diagonal line.

 \vspace{2mm}  
   \noindent\begin{minipage}{0.4\textwidth}
 
{ { Figure 2.  LIGO restrictions on the rate of BH mergers allowed in AGN. Disk scale height and lifetime must live below the upper diagonal line. We assumed $N_{\rm sBH}=2 \times 10^{4}$ \citep{Antonini14,Haileyetal18} and $f_{\rm AGN}=0.3$, corresponding to all LINERs and low luminosity AGNs \citep[LLAGN;][]{Ho2008}. The LIGO-Virgo upper and lower limits are given by the diagonal lines. If other methods or messengers can eliminate the possibility that AGN driven mergers contribute significantly to the LIGO measured rate, we restrict AGN disk parameters to lie below the lower diagonal line. \label{fig:LINERSnotRIAFs}
 }
}
\end{minipage}%
\hfill%
\noindent\begin{minipage}{0.7\textwidth} 
\includegraphics[width=0.8\textwidth]{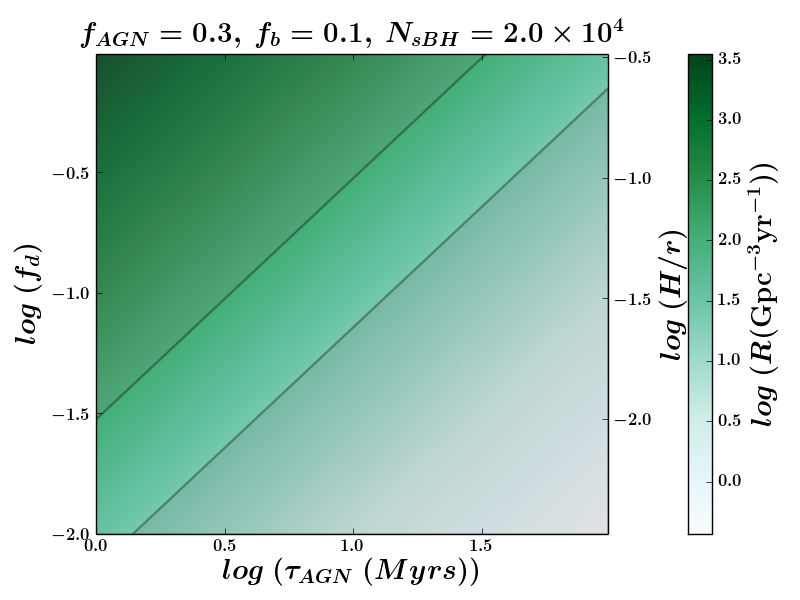}
\vspace{2mm}
\end{minipage}

Thus, if we can predict the spin, mass, and redshift distributions expected from AGN-driven sBBH mergers, we can use the measured LIGO distributions of these parameters to drive the constraints on $h$ and $\tau_{\rm AGN}$ towards the bottom RHS of Fig.~2. {\textbf{Theoretical work on, and simulations of, the behaviour of embedded objects in AGN is key to getting the most out of the growing LIGO distributions of mergers.}} Modelling of the effects of gas-driven inspiral is also important for planned space-based detectors like LISA, which will detect inspiraling sBBHs earlier in their evolution at wider separation only under the assumption of GW-dominated inspiral. {\textbf{Gas-driven mergers could proceed through the LISA frequency range too quickly ($O(\sim \rm{hrs})$) to permit LISA detections of sBBH.}}

\section{What does a mass upper limit from LIGO tell us about AGN?}

If the AGN channel is efficient, then modest-scale IMBHs must grow in all AGN disks \citep{McKF12,McKF14}, particularly at migration traps in those AGN disks \citep{Bellovary16,Secunda18}. Thus upper limits to masses involved in mergers from LIGO put strong limits on disk structure and lifetime. IMBHs can hope to be identified in AGN disks in EM via the radial velocity motions of the inner AGN disk as the SMBH orbits the center of mass of the IMBH-SMBH binary \citep{McKFetal13,McKFord15}. The same technique, searching for systematic perturbations to the broad Fe K$\alpha$ line component, or the blurred lines in the soft X-ray excess, allow us to test the occurrence of extreme mass ratio inspirals (EMRIs) in AGNs, as could a study of ultraviolet variability from the innermost AGN disk. A long-term study of broad Fe K$\alpha$ line variability and soft X-ray excess variability across a large sample of AGNs with present and future missions will provide EM constraints on binarity and EMRIs on a useful sample of AGNs in the relatively local Universe. Such constraints will allow us to put limits on both the efficiency of the sBBH merger process in general, and the population of embedded objects in AGN disks.

\section{What do we need?}
\subsection{A complete catalog of AGNs in the LIGO GW search volume}
If we are to use EM to identify the AGN component of the LIGO mergers, a key 
    requirement
is a complete catalog of AGNs, with little contamination \citep{2017A&ARv..25....2P}, 
    covering 
the LIGO search volume. 

Our goal is to distinguish quiescent galactic nuclei from AGN with a sufficiently massive ($\sim 0.01M_{SMBH}$) gas disk that the dynamics of stars and stellar mass BHs in orbit around the central SMBH are substantially altered. Current all-sky surveys do not provide 
    the required
completeness in any waveband out to the LIGO detection horizon. The most cost-effective strategy for determining the AGN fraction in any given LIGO error box must be dedicated multi-wavelength follow-up on each LIGO error volume; however, follow-up on the best localized events using a complete catalog will provide the most `bang for the buck'.

{\textbf{Mid-IR}} observations are most useful for assembling 
    sufficiently complete
    catalogs of high luminosity AGNs
\citet{Mateos2012}. The WISE color AGN-selection technique is estimated to be $\sim 78\%$ complete at $\sim 95\%$ non-contamination \citep{Stern2012}. AGN catalogs can be produced for nearly arbitrary completeness or contamination fractions (though we must trade off between them) \citep{Assef2018}. Using \citet{Stern2012} completeness and contamination fractions, and assuming the brightest Seyfert galaxies as sBBH merger sites (for an intrinsic $n_{\rm AGN} = 10^{-4}$\,Mpc$^{-3}$) we require
$N_{\rm GW}$ of about 200 to demonstrate a clear link to AGNs, assuming all sBBH mergers come from AGNs. By improving galaxy survey completeness and contamination, we can reduce the needed $N_{\rm GW}$ by nearly a factor of 2.

\subsection{Theoretical work on embedded objects in AGN disks}
If we wish to probe beneath the AGN photosphere using LIGO and LISA, we need:
\begin{itemize}
{\item predictions for distributions of GW parameters \& EM signatures from sBBH mergers in AGN disks.}
{\item hydrodynamic models of multiple objects interacting with the AGN disk, including objects not initially co-planar with the gas.} 
{\item better models of the inspiral of a binary embedded in a differentially rotating disk (there are already many models of binaries in circumbinary disks; however a binary whose center of mass is orbiting a central SMBH may have a substantially different evolution).}
{\item  
    Understanding of the effect of radiation feedback
from accretion onto sBBHs on the surrounding gas and its effect on gas torques for migration and hardening.}
{\item A 
treatment of the effect of tertiary encounters in the AGN disk, including retrograde orbiters and the spherical population interacting with the disk.}
{\item 
     Calculations of
GW waveforms for gas-driven inspiral from the LISA band into the LIGO band. This task is in the LISA Science Consortium plan of work.}
\end{itemize}
Much analogous work has been done for protoplanetary disks, where embedded objects are 
    predominantly
co-planar and on prograde orbits with no additional spherical population.  All of these theoretical studies need to yield distributions of mass ratios  
    $q$ and spins
$\chi_{\rm effective}$ in order for LIGO and LISA to constrain AGN disks most effectively.

\subsection{Multi-wavelength \& multi-messenger constraints on AGNs \& follow-up}
{\textbf{Radio}} follow-up observations of galaxies in GW error regions, combined with optical/mid-IR data, can separate normal star-forming galaxies from AGNs \citep[e.g.][]{Condon1992,Sadler1999,Smolcic2008,Baran2016}. At flux densities $\lesssim 200$\,$\mu$Jy at 3\,GHz, star-forming galaxies dominate the fractional contribution to the total source sample \citep{Baran2016} although LLAGNs may also be present \citep{Mooley2013}. From \S\ref{sec:inference}, we may be able to ignore LLAGNs if they are in normal galaxies. Dwarf galaxy AGNs, however, may host a substantial NSC and a dense gas disk; thus dwarf galaxies must be accounted for in catalogs. The transient variable radio sky at mJys is dominated by AGNs \citep{Sadler1999}. A combination of optical properties, radio spectral index, luminosity, and mid-IR properties can be used to distinguish between, e.g., variable AGN radio emission and supernovae \citep{Mooley2013}.

Most of the {\textbf{cm radio}} emission from normal galaxies is due to synchrotron radiation from SN afterglows; a smaller contribution arises from free-free scattering. By comparing estimates of star formation rates derived using 1.4\,GHz observations with optical constraints, AGN excess ratio emission can be found \citep{Palliyaguru2016}. To separate star-forming galaxies from AGN, radio data must be combined with optical/mid-IR observations \citep{Mauch2007}. The census of radio properties of galaxies provided by VLASS will be an excellent starting point for characterizing galaxies in LIGO error volumes. Future radio observatories (SKA and NGVLA)  will probe the radio sky with deeper sensitivity.
Incompleteness and contamination will impact  follow-up searches. BPT optical line diagnostics can be used to help distinguish classes of activity (LINERs and AGN) as well as SF. So, {\textbf{optical}} line diagnostics are important co-requirements of any catalog.  

{\textbf{UV followup is critically important despite the absence of a major UV mission after the end of the Hubble Space Telescope mission.}} Space-based {\textbf{UV detectors}} with spectroscopic capabilities in the FUV (100--200~nm) and NUV (200--400~nm) will allow us to probe short timescale spectral variability of the innermost AGN disk. Such variability probes disk physics, but also the presence of secondaries or EMRIs. A large field of view permits a rapid search for short-timescale UV variability across many sources.  Modest spectral resolution ($R>$100) allows us to study H$\alpha$ and {Mg\sc{ii}} variability in response to disk changes. An effective area of several $m^{2}$ in FUV/NUV would permit the detection and characterization of short-timescale UV variability in hundreds of sources over a range of redshifts ($z<0.6$) at $5\sigma$ significance in $10$hrs \citep{luv18}. The statistics of short-timescale or periodic variability in the UV-band across a large sample of AGN will allow us to constrain the population of IMBH secondaries.

We need better observational constraints on $N_{sBH}$ in NSCs, especially in our own galactic nucleus. $N_{sBH}$ in our own galactic nucleus is inferred from {\textbf{X-ray binary studies}} \citep{Haileyetal18,Generozov18}, and tests of this inference as well as ongoing, deeper X-ray studies confirming 
    that the X-ray sources are BHs 
rather than 
     neutron stars or white dwarfs
are to be encouraged. {\textbf{X-ray detectors}} with energy resolution of $\sim 2$--6~eV in the Fe K$\alpha$ band permit radial velocity studies of Fe K$\alpha$ emitted by SMBH to detect or limit the existence of smaller mass companions down to mass ratios of $q \sim 0.01$ \cite{McKFetal13,McKFord15}. An effective area of 0.25$m^{2}$ at 6~keV allows the detailed study of the broad Fe K$\alpha$ line in only 30 nearby SMBH \footnote{https://www.cosmos.esa.int/documents/400752/400864/Athena+Mission+Proposal/}. An order of magnitude increase of effective area at 6~keV to $\geq 2.5$~m$^{2}$ is required to establish a statistically significant sample of broad Fe K$\alpha$ lines around SMBHs for study. {\textbf{Work should be carried out to see if soft X-ray blurred lines or soft excess can be used for analogous radial velocity studies}}. 

Searches for common sources of GWs and {\textbf{high-energy neutrinos}} (HEN) are ongoing and potentially will revolutionize our understanding of multi-messenger astrophysical sources \citep[e.g.][]{2018arXiv181010693A}.
The low-latency capability of this search~\citep{2019arXiv190105486C,2017ApJ...850L..35A} will also provide better localization which makes EM follow-up observations of the GWs more efficient. Currently, GW+HEN searches use neutrino data from IceCube, ANTARES, and Pierre Auger observatories and GW data from LIGO and Virgo. 
The next generation of neutrino detectors (IceCube-Gen2 and KM3Net) will increase the sensitivity and discovery potential of these searches by improving the statistics and  angular resolution. 

\section{Summary}
The sBBH mergers presently detected by LIGO may originate wholly or in part from binary BH mergers embedded in disks of gas around SMBHs. Determining the contribution of these AGN disks to the sBBH merger rate enables us to uniquely measure important parameters of AGN disks, including their typical density, aspect ratio, and lifetime, thereby putting unique limits on an important element of galaxy formation. For the first time, GW will allow us to reveal the properties of the hidden interior of AGN disks, while EM probes the disk  photosphere.
The localization of sBBH merger events from LIGO is generally insufficient for association with a single EM counterpart. However, the contribution to the LIGO event rate from rare source types (such as AGNs) can be determined on a statistical basis. To determine the contribution to the sBBH rate from AGNs in the next decade  requires: {\textbf{1) a complete galaxy catalog for the LIGO search volume, 2) strategic multi-wavelength EM follow-up of LIGO events and 3) significant advances in theoretical work on embedded objects in AGN disks and the disks themselves for LIGO and LISA. }}

%
%
%
%

\bibliography{ms.bib}

\end{document}